\documentclass[conference,letterpaper]{IEEEtran}

\usepackage[utf8]{inputenc} 
\usepackage[T1]{fontenc}
\usepackage{url}
\usepackage{ifthen}
\usepackage{cite}
\usepackage[cmex10]{amsmath} 
\usepackage{algorithmic}
\usepackage{algorithm}
\usepackage{graphicx}
\usepackage{cite}
\usepackage{array}
\usepackage{amsmath,amsfonts}
\usepackage{amssymb}
\usepackage{microtype}
\usepackage{comment}

\newtheorem{theorem}{Theorem}
\newtheorem{definition}{Definition}   
\newtheorem{lemma}{Lemma}      
\newtheorem{corollary}{Corollary}    
\newtheorem{remark}{Remark}

\def\BibTeX{{\rm B\kern-.05em{\sc i\kern-.025em b}\kern-.08em
    T\kern-.1667em\lower.7ex\hbox{E}\kern-.125emX}}

\IEEEoverridecommandlockouts
\begin{document}
\title{ Bounds and Constructions of High-Memory Spatially-Coupled Codes 
\thanks{This work is partially supported by the National Key R\&D Program of China, (2023YFA1009600).}}

 \author{
 \IEEEauthorblockN{Lei Huang}
 \IEEEauthorblockA{Data Science Institute\\
                    Shandong University\\
                    Jinan, China\\
                    Email: leihuang@mail.sdu.edu.cn}}

\maketitle

\begin{abstract}
In this paper, we apply the Clique Lovász Local Lemma  to provide sufficient conditions on memory and lifting degree for removing certain harmful combinatorial structures in spatially-coupled (SC) codes that negatively impact decoding performance. Additionally, we present, for the first time, a constructive algorithm based on the Moser-Tardos algorithm that ensures predictable performance. Furthermore, leveraging the properties of LLL-distribution and M-T-distribution, we establish the dependencies among the harmful structures during the construction process. We provide upper bounds on the probability change of remaining harmful structures after eliminating some of them. In particular, the elimination of 4-cycles increases the probability of 6-cycles becoming active by at most a factor of $e^{8/3}$.
\end{abstract}

\section{Introduction}
Spatially-coupled low-density parity-check (SC-LDPC) codes have attracted much attention due to their threshold saturation property \cite{r2,r3,r4} and the capacity-achieving performance over general binary memoryless channels \cite{r5}. In addition, the highly regular structure makes it possible to decode using low-latency windowed decoding (WD) \cite{r6}, \cite{r7}.  Research on SC-LDPC codes can be traced back to the convolutional LDPC codes introduced by Felström and Zigangirov \cite{r1}.

In practical data storage systems, the frame error rate (FER) requirement is below $10^{-12}$ \cite{r8,r9,r10}. Therefore, it is crucial to construct SC-LDPC codes with good bit-error rate (BER) performance in both waterfall and error floor regions under belief-propagation (BP) decoders. In the low-SNR region, the performance of SC-LDPC codes is mainly determined by their asymptotic properties (such as decoding threshold) \cite{r5}, and density evolution (DE) techniques can be used to analyze the decoding threshold of the code. In the high-SNR region, the performance of SC-LDPC codes is determined by finite-length characteristics associated with harmful topological structures  (e.g., cycles and trapping sets) \cite{r9}, \cite{r10}.

Generally speaking, the probability of cycles being active after edge-spreading in the base graph corresponding to matrix H decreases with the increase of memory $m$. When uniform edge-spreading is used, the probability of cycles being active is $O(\frac{1}{m})$, and this result is independent of the length of cycles \cite{r11}. Larger memory makes the graph sparser when edge-spreading reduces the number of harmful structures such as cycles and trapping sets, and it also provides a wider design space. Currently, considerable research focuses on constructing SC-LDPC codes with large girth and minimal harmful structures. Algebraic constructions can accurately characterize the girth and the number of harmful structures under some restrictive conditions, while random constructions can reduce the number of harmful structures by heuristic algorithms and pruning of search space under more general settings. In \cite{r12}, they proposed a combinatorial framework to develop optimal quasi-cyclic (QC) SC codes, comprising so-called optimal overlap (OO) to search for the optimal partitioning matrices, and lifting optimization (CPO) to optimize the lifting matrices, which was later extended by \cite{r13}. In \cite{r14}, they established a probabilistic optimization framework to randomly construct high-memory SC-LDPC codes containing as few harmful structures as possible by using the  locally optimal edge density obtained by gradient descent as the initial condition for heuristic search. This framework was combined with the OO-CPO algorithm to construct codes with good performance in error floor region. However, the uncertainty in the running time of the OO-CPO algorithm poses significant challenges for constructing SC-LDPC codes. 

In this paper, we consider the non-uniform protograph-based SC-LDPC codes in the same probability metric as \cite{r14}, and provide mathematical support for the construction of SC-LDPC codes through  probabilistic combinatorial methods. In the extended version~\cite{r28}, we provide complete proofs of this work, along with further discussion of the new bounds and constructions.

\section{PRELIMINARIES}

\subsection{QC-SC-LDPC Codes}
In this section, we introduce the construction process of QC-SC-LDPC codes using QC-LDPC codes as  underlying block codes. The QC structure not only enables efficient encoding and decoding implementation to SC-LDPC codes \cite{r15}, \cite{r16}, but also makes the constructed codes exhibit anytime-reliable properties \cite{r17}, \cite{r18}.

We consider the QC-SC-LDPC code of Type-I. First, we start with an all-one $(\gamma,\kappa)$  base matrix $\mathbf{H}$, which corresponds to a fully connected bipartite base graph. As the name suggests, SC-LDPC codes are generally constructed by coupling a series of identical block codes into a chain. This operation has different names in different perspectives, but the most widely intuitive is the matrix perspective (\textit{edge-spreading}), which is seen as dividing the edges of the base matrix $\mathbf{H}$ into $m+1$ component matrices $\{\mathbf{H}_0,..., \mathbf{H}_m\}$ of the same size, such that $\mathbf{H}=\sum_{l=0}^{m}\mathbf{H}_l$, where $m$ is \textit{memory}. We record the exact way in which the edge spreads in the $(\gamma,\kappa)$ partition matrix  $\mathbf{P}$, which means that $\mathbf{P}(i,j)=k,~0\le k\le m,~i\in [\gamma], ~j\in [\kappa]$, if and only if $\mathbf{H}_k(i,j)=1, ~\mathbf{H}_l(i,j)=0, ~\forall l\neq k, ~0\le l\le m$. Generally, the base matrices and edge-spreading patterns may differ among uncoupled codes, like those constructed via the random spreading method for the proof of universality. However, considering the implementation complexity in practical systems, the base matrices and the edge-spreading patterns of all uncoupled codes are assumed to be identical in this paper. By concatenating these component matrices vertically from $\mathbf{H}_0$ to $\mathbf{H}_m$, a \textit{replica} is obtained. By concatenating $L$ of these identical  replicas horizontally in space, we get the protograph matrix $\mathbf{H}_{SC}^P$ of the SC-LDPC code, where $L$ is called the \textit{coupling length}.

To construct QC-SC-LDPC codes, after the edge-spreading process, we perform the lifting operation on the protograph. The lifting for every uncoupled code is identical, which facilitates the practical implementation of the windowed decoder. Let $\mathbf{L}$ be a $\gamma \times \kappa$ matrix called the \textit{lifting matrix}. $\sigma$ denotes the $Z \times Z$ circulant matrix obtained by cyclically shifting the columns of an identity matrix one unit to the left. For all $i \in [\gamma], ~j \in [\kappa]$, if $\mathbf{L}(i,j)=x$ where $x \in \{0,\dots,Z-1\}$, then replace the ones in $\mathbf{H}_{SC}^P$ corresponding to $\mathbf{H}(i,j)$ with the $\sigma^x$ matrix, and replace the remaining zeros with $Z \times Z$ zero matrices. Under this notation, the parameters of a QC-SC-LDPC code $\mathbf{H}_{SC}$ are described by $(\gamma,\kappa,m,Z,L)$.

\subsection{Major Harmful Structures}

Despite the fact that QC-SC-LDPC codes offer improved performance compared to their underlying block codes, they still experience performance degradation at high signal-to-noise ratios (SNRs) over the additive white Gaussian noise (AWGN) channel under iterative decoding, resulting in an error floor phenomenon. This error floor is primarily caused by small structures in the code’s Tanner graph, known as trapping sets (TSs), which, in turn, consist of small cycles. We review the key definitions:

\begin{lemma}\label{lemma_1}
Let $\mathcal{C}_{2g}$ be the set of cycle-2g candidates in the base matrix, where $g \in \mathbb{N},~g \geq 2$. Denote $c_{2g} \in \mathcal{C}_{2g}$ by $(j_1, i_1, j_2, i_2, \dots, j_g, i_g)$, where $(i_k, j_k)$ and $(i_k, j_{k+1})$ for $1 \leq k \leq g$, with $j_{g+1} = j_1$, are nodes of $c_{2g}$ in $\mathbf{H}$, $\mathbf{P}$, and $\mathbf{L}$. Then $c_{2g}$ becomes a cycle candidate in the protograph after the edge-spreading operation if and only if the following condition holds \cite{r19}:

\begin{align}\label{e_1}
\sum_{k=1}^{g} \mathbf{P}(i_k, j_k) = \sum_{k=1}^{g} \mathbf{P}(i_k, j_{k+1}). 
\end{align}

This cycle candidate in the protograph becomes a cycle in the Tanner graph after the Z-lifting operation if and only if the following condition holds \cite{r20}:

\begin{align}\label{e_2}
 \sum_{k=1}^{g} \mathbf{L}(i_k, j_k) = \sum_{k=1}^{g} \mathbf{L}(i_k, j_{k+1}) \mod Z.   
\end{align}

\end{lemma}

To reduce the complexity of optimizing QC-SC-LDPC codes, the process is typically divided into two stages. First, matrix $\mathbf{P}$ is optimized to minimize cycle candidates in the protograph. This sets the stage for optimizing matrix $\mathbf{L}$, aiming to further reduce cycles in the Tanner graph. This two-step method efficiently streamlines optimization and reduces cycles, as shown in \cite{r12}, \cite{r14}.

In the remainder of this paper, we focus on QC-SC-LDPC codes for the AWGN channel, where the dominant harmful structures are the low-weight absorbing sets (ASs) \cite{r12}. The ASs are defined as follows:

\begin{definition}\label{def}{(Absorbing Sets)}
 Consider a subgraph induced by a subset $\mathcal{S}$ of $V$ (VNs) in the Tanner graph of a code. 
The subset $\mathcal{N}(\mathcal{S})$ of $C$ (CNs) denotes the set of neighbors of $\mathcal{S}$. $\mathcal{N}(\mathcal{S})=\mathcal{N}_o(\mathcal{S})\bigcup \mathcal{N}_e(\mathcal{S})$, where $\mathcal{N}_o(\mathcal{S})$ is the set of nodes with odd degree (unsatisfied), and $\mathcal{N}_e(\mathcal{S})$ is the set of nodes with even degree (satisfied).
The set $\mathcal{S}$ is said to be an $(a, b)$  \textit{absorbing set} (AS)  if the size of $\mathcal{S}$ is $a$, the number of unsatisfied neighboring CNs of $\mathcal{S}$ is $b$ and all the variable nodes in $\mathcal{S}$ are connected to more satisfied neighboring CNs than unsatisfied neighboring CNs. 

\end{definition}

\subsection{Lovász Local Lemma}

The Lovász Local Lemma (LLL), introduced by Erdős and Lovász in 1975, has become one of the most important probabilistic methods. LLL provides sufficient conditions under which a set of undesirable events $\mathcal{A}$  in a probability space $\Omega$ can be simultaneously avoided, meaning the conditions for $P(\cap_{A\in \mathcal{A}}\Bar{A})>0$ to hold. The relationships between these bad events are depicted by an undirected graph called the \textit{dependency graph} $G_D=([n],E)$ with $|G|:=n$ and $\|G\|:=|E|$, where each bad event corresponds to a vertex in the dependency graph. An event $A_i$ is independent of $\{A_j:j\neq i,j\notin \mathcal{N}(i)\}$, where $\mathcal{N}(i)$ stands for the neighborhood of $A_i$ in $G_D$. The \emph{scope} of an event $A$, denoted $sc(A)$, is the minimal set of random variables on which $A$ depends. In what follows, we introduce an enhanced version of LLL, leveraging additional local structural information.

\begin{lemma}\label{lemma_2}{(Quantitative Clique Lovász Local Lemma \cite{r21})}
Let $\mathcal{A}=\{A_1,A_2,\ldots,A_n\}$ be a set of events with dependency graph $G_D$ and let $\mathcal{K}=\{K_1,K_2,\ldots,K_c\}$ be a set of cliques in $G_D$ covering all the edges (not necessarily disjointly). If there exists a set of vectors $\{\textbf{x}_1,\textbf{x}_2,\ldots,\textbf{x}_c\}$ from $(0,1)^n$ such that following conditions are satisfied:
\begin{enumerate}
    \item $\forall v \in [c]:~\sum_{i\in K_v}x_{i,v}<1$
    \item $\forall i \in [n],~\forall v$ such that $i\in K_v$:
    \begin{equation}
        P(A_i)\le x_{i,v}\prod_{u\neq v:i\in K_u}(1-\sum_{j\in K_u \backslash \{i\}}x_{j,u})
    \end{equation}
\end{enumerate}
then the following hold:
\begin{enumerate}
    \item $P(\bigcap_{i\in [n]}\Bar{A_i})\ge \prod_{v\in [c]}(1-\sum_{i\in K_v}x_{i,v})>0$
    \item In the variable framework, the running time of the Moser–Tardos algorithm ( Algorithm~\ref{alg_1} ) is at most:
    \begin{equation}
        \sum_{i\in [n]}min_{v:i\in K_v}\frac{x_{i,v}}{1-\sum_{j\in K_v}x_{j,v}}
    \end{equation}
\end{enumerate}
\end{lemma}

\section{Upper Bounds and Constructions}

\subsection{Upper Bounds for Lifting Degree and Memory}

In this section, we present sufficient conditions for constructing QC-SC-LDPC codes that eliminate specific harmful structures. These conditions correspond to upper bounds on the lifting degree and memory.

For convenience, we first define some essential parameters related to edges in the base graph, which will come up frequently in subsequent calculations and discussions.

\begin{definition}For any QC-SC-LDPC code constructed with respect to a $\gamma \times \kappa$ fully-connected block code base graph $G_{B}=(V,E)$, with no parallel edges. The \textit{harmful weight factor} $W_{\mathcal{H}}^e$  of an edge $e$ in the base graph with respect to the harmful structure set $\mathcal{H}=\{H_1,...,H_k\}$ is the number of harmful structures passing through the edge in the base graph that belong to the above set. Let $W_{\mathcal{H}} :=\max\{W_{\mathcal{H}}^e|e\in E(G_{B})\}$  be the \textit{maximum harmful weight factor}.
\end{definition}

\begin{remark}
Since the size of the base graph is usually quite small, determining this parameter is not difficult.
\end{remark}

\addtolength{\topmargin}{0.02in}

\begin{lemma}\label{lemma_3}
Consider a random cyclic $Z$-lifting of a base bipartite graph $G$ with no parallel edges, and consider a harmful structure $H$ with $n_b$ fundamental cycles $\{c_1,\dots,c_{n_b}\}$. Let $P_H^l$ denote the probability that $H$ is active after a random lifting. We then have
\begin{align}
 P_H^l \leq \frac{\prod_{i=1}^{n_b}|c_i|}{(4Z)^{n_b}}.
\end{align}
\end{lemma}
\begin{IEEEproof}
    The proof is obtained by combining \cite[Theorem 2]{r22} with \cite[Lemma 3]{r23}.
\end{IEEEproof}

\begin{remark}
    The probability that $H$ is active after random edge-spreading, denoted by $P_H^s$, was derived in \cite{r14}.
\end{remark}

\begin{theorem}\label{theorem_1}
 For any QC-SC-LDPC code constructed with respect to a $\gamma \times \kappa$ fully-connected block code base graph, with no parallel edges, under the coupling pattern $\mathbf{a}=(a_0,a_1,\ldots,a_{m_t})$ and corresponding probability distribution $\mathbf{p}=(p_0,p_1,\ldots,p_{m_t})$. $\mathcal{H}=\{H_1,\ldots,H_k\}$ are a series of avoidable harmful structures in the base graph with dependency graph $G$, and $\{P_{H_1},\ldots,P_{H_k}\}=\{P(H_1),\ldots,P(H_k)\}$ are the probabilities that they are active after the random edge-spreading and lifting process, if
\begin{align}
    \max\{P_{H_1},\ldots,P_{H_k}\} \le \max\{\underbrace{\frac{(\Delta-1)^{\Delta-1}}{(\Delta)^{\Delta}}}_{I},\underbrace{\frac{(|H| - 1)^{|H| - 1}}{(W_\mathcal{H} - 1)|H|^{|H|}}}_{II} \}
    \nonumber 
\end{align}
Here, $\Delta$ is the maximum degree of the dependency graph, and $|H| = \max\{|H_1|, \ldots, |H_k|\}$ is the maximum number of vertices in the harmful structures.
Then
\begin{align}
{P(\bigcap_{i\in [k]}\Bar{H_i})} \ge \begin{cases}
(1-\frac{2}{\Delta})^{\|G\|},&{\text{if}}\ I>II \\ 
{(1-\frac{W_{\mathcal{H}}}{(W_{\mathcal{H}}-1)|H|})^{\gamma\kappa},}&{\text{otherwise.}} 
\end{cases}
\end{align}

\end{theorem}

\begin{IEEEproof}
\begin{itemize}

    \item Bad Events:\\
    Each avoidable harmful structure $H_i$ in the base graph is an event, which is a constraint imposed on the set of random variables corresponding to the edges passed through, when the value of the random variable is taken such that this structure is active, it corresponds to the occurrence of this bad event. 
    \item Probability of bad events:\\
    The probability that avoidable harmful structures $H_i$ is active after the random edge-spreading and lifting process is $P(H_i)=P_{H_i}^l\times P_{H_i}^s $.

    \item Construction of cliques:\\
     Take each event as a vertex and if two different events involve at least one of the same random variable, connect them and we get the dependency graph $G$.
Now, we consider any two vertices in $G$ that are connected to each other as a clique, so that we find the set $\{K_1,K_2,...,K_c\}$ of such cliques that satisfy the lemma, where $c=\|G\|$.
\end{itemize}

Assuming $\forall v \in [c], i \in [k]$, $x_{i,v} = x$, the conditions in Lemma \ref{lemma_2} reduces to:

\begin{enumerate}
    \item $2x<1$
    \item 
        \begin{align}
        P(H_i)\le x(1-x)^{\Delta-1}
         \end{align}
\end{enumerate}
To satisfy condition (1) while maximizing the right-hand side of inequality in condition (2), let
    \begin{align}
      x=\frac{1}{\Delta}
    \end{align}
Then, condition (2) reduces to:
\begin{align}
    P(H_i)\le \frac{(\Delta-1)^{\Delta-1}}{(\Delta)^{\Delta}}
\end{align}

 By utilizing the structural information of the dependency graph, we reconstruct the set of cliques $\{K_1, K_2,\ldots, K_c\}$ to obtain tighter bounds. Note that for each vertex in the dependency graph, many of its neighbors are interconnected, allowing us to group the vertex's neighbors into fewer, larger cliques. For each edge in the $\gamma \times \kappa$ base graph, the set of harmful structures passing through it forms a clique in the dependency graph. The set of these cliques $\{K_1, K_2, ..., K_{\gamma \kappa}\}$ clearly covers all edges in the dependency graph. In this case, the neighbors of each vertex in the dependency graph can be grouped into $|H|$ cliques. Assuming $\forall v \in [c],~i \in [k]$, $x_{i,v} = x$, the conditions in Lemma \ref{lemma_2} reduces to:
  \begin{enumerate}
      \item $W_{\mathcal{H}}x < 1$
      \item 
      \begin{align}
          P(H_i) &\le x(1 - (W_{\mathcal{H}}-1)x)^{|H| - 1}\\
          & \le x\prod_{u \neq v: i \in K_u} \left(1 - \sum_{j \in K_u \backslash \{i\}} x \right)
      \end{align}     
  \end{enumerate}
To satisfy condition (1) while maximizing the right-hand side of inequality in condition (2), let
\begin{align}
 x = \frac{1}{(W_{\mathcal{H}} - 1)|H|}   
\end{align}
Then, condition (2) reduces to:
\begin{align}
   P(H_i) \le \frac{1}{W_{\mathcal{H}} - 1} \frac{(|H| - 1)^{|H| - 1}}{|H|^{|H|}} 
\end{align}

Note that
\begin{align}
     \frac{(\Delta-1)^{\Delta-1}}{(\Delta)^{\Delta}}&=\frac{1}{\Delta}\frac{(\Delta-1)^{\Delta-1}}{(\Delta)^{\Delta-1}}\\
     \frac{1}{W_{\mathcal{H}} - 1} \frac{(|H| - 1)^{|H| - 1}}{|H|^{|H|}}&=\frac{1}{(W_{\mathcal{H}} - 1)|H|} \frac{(|H| - 1)^{|H| - 1}}{|H|^{|H|-1}} 
\end{align}

The function $\left(1 - \frac{1}{x}\right)^{x - 1}$ is decreasing, and
\begin{align}
    \lim_{x \to +\infty} \left(1 - \frac{1}{x}\right)^{x - 1} = \frac{1}{e}
\end{align}
Since $\Delta > |H|$ and $(W_{\mathcal{H}} - 1)|H| > \Delta$ always hold, we obtain:
\begin{align}
    \frac{(|H| - 1)^{|H| - 1}}{|H|^{|H|-1}}& >\frac{(\Delta-1)^{\Delta-1}}{(\Delta)^{\Delta-1}}\\
    \frac{1}{\Delta}&>\frac{1}{(W_{\mathcal{H}} - 1)|H|}
\end{align}

Therefore, we can obtain the best possible bound by:

\begin{align}
  P(H_i)  \le \max\left\{\frac{(\Delta - 1)^{\Delta - 1}}{\Delta^{\Delta}}, \frac{(|H| - 1)^{|H| - 1}}{(W_\mathcal{H}-1)|H|^{|H|}}\right\}  
\end{align}

Based on Lemma \ref{lemma_2}, we know that when:
\begin{align}
    \max\{P_{H_1},\ldots,P_{H_k}\} \le \max\{\underbrace{\frac{(\Delta-1)^{\Delta-1}}{(\Delta)^{\Delta}}}_{I},\underbrace{\frac{(|H| - 1)^{|H| - 1}}{(W_\mathcal{H} - 1)|H|^{|H|}}}_{II} \}
    \nonumber 
\end{align}
Conclusion (1) in Lemma \ref{lemma_2} holds, that is:

\begin{align}
{P(\bigcap_{i\in [k]}\Bar{H_i})} \ge \begin{cases}
(1-\frac{2}{\Delta})^{||G||},&{\text{if}}\ I>II \\ 
{(1-\frac{W_{\mathcal{H}}}{(W_{\mathcal{H}}-1)|H|})^{\gamma\kappa},}&{\text{otherwise.}} 
\end{cases}
\end{align}
Hence, matrices $\mathbf{P}$ and $\mathbf{L}$ exist such that all harmful structures are avoided.
    
\end{IEEEproof}
\addtolength{\topmargin}{0.011in}

\begin{lemma}{(\cite{r11})}\label{lemma 4}
 Let $G$ be a base graph of a block code and suppose that we have a TBC walk $c_4\in\mathcal{C}_4$ of length 4 in that base graph. If we consider memory $m$, coupling length $L\ge m+1$ and apply the edge spreading process under the coupling pattern $\mathbf{a}=(0,1,\ldots,m)$ with  uniform probability distribution $\mathbf{p}=\frac{1}{m+1}\mathbf{1}_{m+1}$, then the probability that $c_4$ remains in the SC base graph is given by
\begin{align}{P^s_{c_4}(\mathbf{a},\mathbf{p})} = \frac{{2{m^2} + 4m + 3}}{{3{{\left( {m + 1} \right)}^3}}}.\end{align}
\end{lemma}

\begin{corollary}\label{corollary 1}
For any QC-SC-LDPC code constructed with respect to a $\gamma \times \kappa$ fully-connected block code base graph, with no parallel edges, a sufficient condition for having a girth of at least 6 is 
\begin{align}
    \frac{2m^2+4m+3}{3(m+1)^3Z} \le \max\{\underbrace{\frac{(\Delta-1)^{\Delta-1}}{(\Delta)^{\Delta}}}_{I},\underbrace{\frac{27}{256(\gamma \kappa -\gamma -\kappa)}}_{II} \}
\end{align}
where $\Delta=(2\gamma-3)(2\kappa-3)$, $Z$ is the lifting degree.
And
\begin{align}
{P(\bigcap_{c_4\in \mathcal{C}_4}\Bar{c_4})} \ge \begin{cases}
(1-\frac{2}{\Delta})^{\frac{\gamma(\gamma-1)\kappa(\kappa-1)\Delta}{8}},&{\text{if}}\ I>II \\ 
{(1-\frac{\gamma \kappa-\gamma-\kappa+1}{4(\gamma \kappa-\gamma-\kappa)})^{\gamma\kappa},}&{\text{otherwise.}} 
\end{cases}
\end{align}

\end{corollary}

\begin{IEEEproof}
    When $\mathcal{H}=\mathcal{C}_4$, $|H|=4$, $|G|=\binom{\gamma}{2}\binom{\kappa}{2}$, $||G||\le \frac{|G|\Delta}{2}$, $W_{\mathcal{C}_4}=\binom{\gamma-1}{1}\binom{\kappa-1}{1}$.
\end{IEEEproof}

Benefiting from the utilization of additional local  information about the dependency graph as described in Lemma \ref{lemma_2},  we  slightly improve the bound presented in \cite{r26}.

\subsection{Theoretical Analysis of Algorithms}
In this section, we present a constructive polynomial algorithm within the framework of the LLL  and provide a theoretical analysis of the algorithm. 
\begin{theorem}\label{theorem_2}
If the conditions in Theorem~\ref{theorem_1} are satisfied, then Algorithm~\ref{alg_1} successfully constructs QC-SC-LDPC codes free from the specified harmful structures in polynomial expected time. More precisely, the expected number of RESAMPLE calls in the  Algorithm~\ref{alg_1} is bounded by:
\begin{equation*}
\sum_{H_i\in \mathcal{H}}{\mathbb{E}}[H_i]\le \begin{cases}
\frac{k}{\Delta-2},&{\text{if}}\ I>II \\ 
{\frac{k}{(W_{\mathcal{H}}-1)|H|-W_{\mathcal{H}}},}&{\text{otherwise.}} 
\end{cases}
\end{equation*}
\emph{(Here $\mathbb{E}[H_i]$ denotes the expected number of times event $H_i$
is resampled by the RESAMPLE procedure in Algorithm~\ref{alg_1}.)}
\end{theorem}

\begin{IEEEproof}
    It directly follows from Lemma \ref{lemma_2}.
\end{IEEEproof}

\begin{corollary}
If the conditions in Corollary~\ref{corollary 1} are satisfied, then Algorithm~\ref{alg_1} successfully constructs QC-SC-LDPC codes free from the specified harmful structures in polynomial expected time. More precisely, the expected number of RESAMPLE calls in the  Algorithm~\ref{alg_1} is bounded by:
\begin{equation*}
\sum_{c_i\in \mathcal{C}_4}{\mathbb{E}}[c_i] \le \begin{cases}
\frac{\binom{\gamma}{2}\binom{\kappa}{2}}{(2\gamma-3)(2\kappa-3)-2},&{\text{if}}\ I>II \\ 
{\frac{\binom{\gamma}{2}\binom{\kappa}{2}}{3(\gamma\kappa-\gamma-\kappa)-1},}&{\text{otherwise.}} 
\end{cases}
\end{equation*}
\emph{(Here $\mathbb{E}[c_i]$ denotes the expected number of times event $c_i$
is resampled by the RESAMPLE procedure in Algorithm~\ref{alg_1}.)}
\end{corollary}

\begin{IEEEproof}
    It directly follows from Lemma \ref{lemma_2}.
\end{IEEEproof}

\begin{figure}[!ht]
		\label{f2}
		\renewcommand{\algorithmicrequire}{\textbf{Input:}}
		\renewcommand{\algorithmicensure}{\textbf{Output:}}
		\begin{algorithm}[H]
			\caption{The Moser-Tardos (MT) Algorithm \cite{r24}}
			\begin{algorithmic}[1]\label{alg_1}
				\REQUIRE  
                          Sample space: $\mathbf{a}=(a_0,a_1,\ldots,a_{m_t})$.\\
                          Independent r.v. taking values in $\mathbf{a}$: $\chi=\{X_1,...,X_{\gamma\kappa}\}$.\\
                          Set of (ordered) events: $\mathcal{H}=\{H_1,...,H_k\}.$\\
                          Probability distribution: $\mathbf{p}=(p_0,p_1,\ldots,p_{m_t})$.
				\ENSURE   Assignment $\alpha =(r_1,...,r_{\gamma\kappa})\in \mathbf{a}^{\gamma\kappa}$ to  r.v.   
                          $\chi$  s.t. $\cap_{H_i \in \mathcal{H}}\Bar{H_i}=TRUE$.
				\STATE Sample the variables $X_i,~i\in[\gamma\kappa]$, and let $\alpha$ be the resulting assignment.
				\WHILE {there exists a bad event in $\mathcal{H}$ that occurs under the current assignment, let $H_j$ be the least indexed such event}
				\STATE RESAMPLE($H_j$)
				\ENDWHILE
                \STATE Output current assignment $\alpha$.
			\end{algorithmic}
            RESAMPLE($H_j$)
            \begin{algorithmic}[1]
                    \STATE Resample the variables in sc($H_j$).
                    \WHILE {there is a least indexed bad event $H_l$, such that sc($H_j$) $\cap$ sc($H_l$) $\neq \emptyset$, occurring under the current assignment,}
                    \STATE RESAMPLE($H_l$)
                    \ENDWHILE
            \end{algorithmic}
		\end{algorithm}
	\end{figure}

\section{ Characterization of the Solution Space}
In this section, we characterize the solution space by studying the changes in its probability distribution under LLL conditions. We prove that during this process, harmful structures associated with more eliminated harmful structures are more likely to remain active, and we provide an upper bound for these changes. Specifically, eliminating short cycles tends to increase the presence of longer cycles. This is consistent with the phenomena observed in previous experimental works.

We first introduce two concepts from \cite{r25} that describe the probability distribution under LLL conditions and the output of the MT algorithm.
\begin{definition}
  The distribution of $\Omega$ conditioned on avoiding $\mathcal{A}$ is called the LLL-distribution. The distribution at the MT algorithm termination is called the MT-distribution.  
\end{definition}

\begin{remark}
    In this section, due to content limitations, we directly utilize the results under the LLL condition. The corresponding results for CLLL will be provided in the extended version of this paper.
\end{remark}

The LLL is a probabilistic tool to generate combinatorial structures with good local properties. The LLL-distribution further shows that these structures have good global properties in expectation. Thus, in a certain sense, the LLL-distribution is a mildly distorted version of the space $\Omega$. For the asymmetric LLL, we have the following bound:

\addtolength{\topmargin}{0.001in}
\begin{lemma}{(\cite{r26})}\label{lemma 5}
If the conditions in  asymmetric LLL\cite[Theorem 1.1]{r27} are satisfied, then the LLL-distribution $D_{LLL}$ is well-defined. For any event $E$ determined by the set of random variables $\chi$, the probability $P_{LLL}[E]$ under the LLL-distribution $D_{LLL}$ satisfies:
\begin{align}
    P_{LLL}[E]:=P[E|\bigwedge_{A\in \mathcal{A} }\bar{A}]\le P_{\Omega}[E]\cdot\prod_{B\in \mathcal{N}(E)}(1-x_B)^{-1}
\end{align}    
Specifically, for symmetric LLL\cite[Theorem 1.5]{r27}, there is a more intuitive result.
\begin{align}
    P_{LLL}[E]:=P[E|\bigwedge_{A\in \mathcal{A} }\Bar{A}]\le P_{\Omega}[E]\cdot(1+ep)^{|\mathcal{N}(E)|}
\end{align}   
\end{lemma}

Moreover, it was shown in~\cite{r25} that the MT distribution $D_{MT}$ also satisfies the above bounds. The lemma above indicates that when an event $E$ is associated with relatively few bad events $A\in \mathcal{A}$, its probability of occurring does not increase significantly.

\begin{theorem}\label{theorem 3}
   For any QC-SC-LDPC code constructed with respect to a $\gamma \times \kappa$ fully-connected block code base graph, with no parallel edges, under the coupling pattern $\mathbf{a}=(a_0,a_1,...,a_{m_t})$ and corresponding probability distribution $\mathbf{p}=(p_0,p_1,...,p_{m_t})$. Suppose $\mathcal{H}=\{H_1,...,H_k\}$ are a series of avoidable harmful structures in the base graph with dependency graph $G$, and $\{P_{H_1},...,P_{H_k}\}$ are the probabilities that they are active after the random edge-spreading and lifting process. If $\mathcal{H}'=\{H_1,...,H_f\}\subseteq \mathcal{H}=\{H_1,...,H_k\}$ with dependency graph $G'= G[\mathcal{H}']$ satisfies the conditions in asymmetric LLL\cite[Theorem 1.1]{r27}, then for any event $E$ determined by the set of random variables $\chi$, the probability $P_{LLL}[E]$ under the LLL-distribution $D_{LLL}$ satisfies:
\begin{align}
    P_{LLL}[E]:=P[E|\bigwedge_{H_i\in \mathcal{H}' }\Bar{H_i}]\le P_{\Omega}[E]\prod_{B\in \mathcal{N}_G(E)\cap \mathcal{H}'}\frac{1}{1-x_B}
\end{align} 
And the probability of $E$ being true in the output distribution of Algorithm \ref{alg_1} also obeys this upper bound.
\end{theorem}

From Theorem \ref{theorem 3}, we learn that more complex harmful structures, due to occupying more edges, are more likely to significantly increase in probability. Specifically, eliminating short cycles makes longer cycles more likely to increase in probability, which is consistent with the experimental observations in \cite{r26}. This suggests that simply increasing the girth may lead to a substantial increase in the number of complex harmful structures, resulting in suboptimal code construction. Therefore, we should balance large girth with a reduced number of complex harmful structures to select the best code under specific conditions.

\begin{corollary}
   For any QC-SC-LDPC code constructed with respect to a $\gamma \times \kappa$ fully-connected block code base graph, with no parallel edges. If $\mathcal{H}'=\mathcal{C}_4$ satisfies the conditions in symmetric LLL\cite[Theorem 1.5]{r27}, then $P_{LLL}[c_{2k}] \le P_{\Omega}[c_{2k}]\cdot(1+\frac{1}{\Delta})^{2kW_{\mathcal{H}'}}$. Specifically,
    $P_{LLL}[c_{6}] \le P_{\Omega}[c_{6}] \cdot e^{8/3}$.

\end{corollary}
\begin{IEEEproof}
According to Lemma \ref{lemma 5}, we have
\begin{align}
 P_{LLL}[c_{2k}] &:=P[c_{2k}|\bigwedge_{c_4\in \mathcal{C}_4 }\bar{c_4}]\le P_{\Omega}[c_{2k}]\cdot(1+ep)^{|\mathcal{N}(E)|}\\
 &\le P_{\Omega}[c_{2k}]\cdot(1+\frac{1}{\Delta})^{|\mathcal{N}(E)|}\\
 &\le P_{\Omega}[c_{2k}]\cdot(1+\frac{1}{\Delta})^{2kW_{\mathcal{H}'}}
\end{align}
For $\mathcal{C}_6$, $\Delta=(2\gamma-3)(2\kappa-3)$, $W_{\mathcal{C}_4}=(\gamma-1)(\kappa-1)$.
\begin{align}
    P_{LLL}[c_{6}]& \le P_{\Omega}[c_{6}]\cdot\left(1+\frac{1}{(2\gamma-3)(2\kappa-3)}\right)^{6(\gamma-1)(\kappa-1)}\\
    &\le P_{\Omega}[c_{6}] \cdot e^{8/3} ~(\gamma,\kappa \ge 3)
\end{align}
\end{IEEEproof}

\IEEEtriggeratref{14}

\bibliographystyle{IEEEtran}

\bibliography{IEEEabrv,ref.bib}

\end{document}